\documentclass[conference]{IEEEtran}
\usepackage[nolist,nohyperlinks]{acronym}
\usepackage{graphicx}
\usepackage{color, xcolor}

\ifCLASSINFOpdf
\else
\fi
\newcommand\jinfeng[1]{{\color{black}#1}}

\begin{document}
%
\title{Decentralized Multi-Party Multi-Network AI for Global Deployment of 6G Wireless Systems \vspace{-1.9em}}



%
\author{\IEEEauthorblockN{Merim Dzaferagic\IEEEauthorrefmark{1},
Marco Ruffini\IEEEauthorrefmark{1},
Nina Slamnik-Krijestorac\IEEEauthorrefmark{7}, 
Joao F. Santos\IEEEauthorrefmark{3},
Johann Marquez-Barja\IEEEauthorrefmark{7},\\
Christos Tranoris\IEEEauthorrefmark{4}, 
Spyros Denazis\IEEEauthorrefmark{4},
Thomas Kyriakakis\IEEEauthorrefmark{12},
Panagiotis Karafotis\IEEEauthorrefmark{12},
Luiz DaSilva\IEEEauthorrefmark{3},\\
Shashi Raj Pandey\IEEEauthorrefmark{5},
Junya Shiraishi\IEEEauthorrefmark{5}, 
Petar Popovski\IEEEauthorrefmark{5},
S{\o}ren Kejser Jensen\IEEEauthorrefmark{5}, 
Christian Thomsen\IEEEauthorrefmark{5}, \\
Torben Bach Pedersen\IEEEauthorrefmark{5}, 
Holger Claussen\IEEEauthorrefmark{6}\IEEEauthorrefmark{2}\IEEEauthorrefmark{1}, 
\jinfeng{Jinfeng Du}\IEEEauthorrefmark{8},
Gil Zussman\IEEEauthorrefmark{9}, 
Tingjun Chen\IEEEauthorrefmark{10}, 
Yiran Chen\IEEEauthorrefmark{10},\\
Seshu Tirupathi\IEEEauthorrefmark{13},
Ivan Seskar\IEEEauthorrefmark{11}, and
Daniel Kilper\IEEEauthorrefmark{1}
}
\IEEEauthorblockA{\IEEEauthorrefmark{1}Trinity College Dublin, Ireland,
\IEEEauthorrefmark{7}University of Antwerp - imec, Belgium, 
\IEEEauthorrefmark{3}Virginia Tech, US, 
\IEEEauthorrefmark{4}University of Patras, Greece, }
\IEEEauthorblockA{\IEEEauthorrefmark{12}Dienekes, Greece, 
\IEEEauthorrefmark{5}Aalborg University, Denmark, 
\IEEEauthorrefmark{6}Tyndall National Institute, Ireland,
\IEEEauthorrefmark{2}University College Cork, Ireland,\\
\IEEEauthorrefmark{8}Nokia Bell Labs, USA,
\IEEEauthorrefmark{9}Columbia University, USA,
\IEEEauthorrefmark{10}Duke University, USA,
\IEEEauthorrefmark{11}Rutgers University, USA,\\
\IEEEauthorrefmark{13}IBM Research Europe}
}



\maketitle

\begin{abstract}
Multiple visions of 6G networks elicit Artificial Intelligence (AI) as a central, native element. When 6G systems are deployed at a large scale, end-to-end AI-based solutions will necessarily have to encompass both the radio and the fiber-optical domain. This paper introduces the Decentralized Multi-Party, Multi-Network AI (DMMAI) framework for integrating AI into 6G networks deployed at scale. DMMAI harmonizes AI-driven controls across diverse network platforms and thus facilitates 
networks that autonomously configure, monitor, and repair themselves. This is particularly crucial at the network edge, where advanced applications meet heightened functionality and security demands. The radio/optical integration 
is vital due to the current compartmentalization of AI research within these domains, which lacks a comprehensive understanding of their interaction. Our approach explores multi-network orchestration and AI control integration, filling a critical gap in standardized frameworks for AI-driven coordination in 6G networks. The DMMAI framework is a step towards a global standard for AI in 6G, aiming to establish reference use cases, data and model management methods, and benchmarking platforms for future AI/ML solutions. \vspace{-1em}
\end{abstract}


%
\IEEEpeerreviewmaketitle

\section{Introduction}

Despite the promise of \ac{ai}, there remain obstacles to its wide adoption in communication networks. For example, the introduction of software-defined elements, such as the O\mbox{-}RAN Alliance's \acp{ric}~\cite{polese2023understanding}, enables the control and management of \acp{ran} by multi-party applications. However, the control and management capabilities of these software-defined elements do not extend to optical networks, and related \ac{ai} functions are still under active research and standardization~\cite{lee2020hosting}.

The high \jinfeng{throughput} and low latency expected for many 6G applications are such that the optical links can no longer be simply viewed as fat pipes, but instead must be managed together with the radio resources for end-to-end efficiency and scale~\cite{filgueiras2023wireless}. \ac{ai} controls can play an important role in managing the diverse and complex requirements of these different network platforms. However, \ac{ai} research is largely siloed within the wireless and optical research communities, and there is little understanding of how \ac{ai}-based controls might interact and jointly manage resources across network domains~\cite{moerman2020mandate,slyne2024demonstration}. Consequently, there does not exist any framework or accepted practice for \ac{ai}-based control and management across radio and optical fibre networks.

\begin{figure}
    \centering
    \includegraphics[scale=0.07]{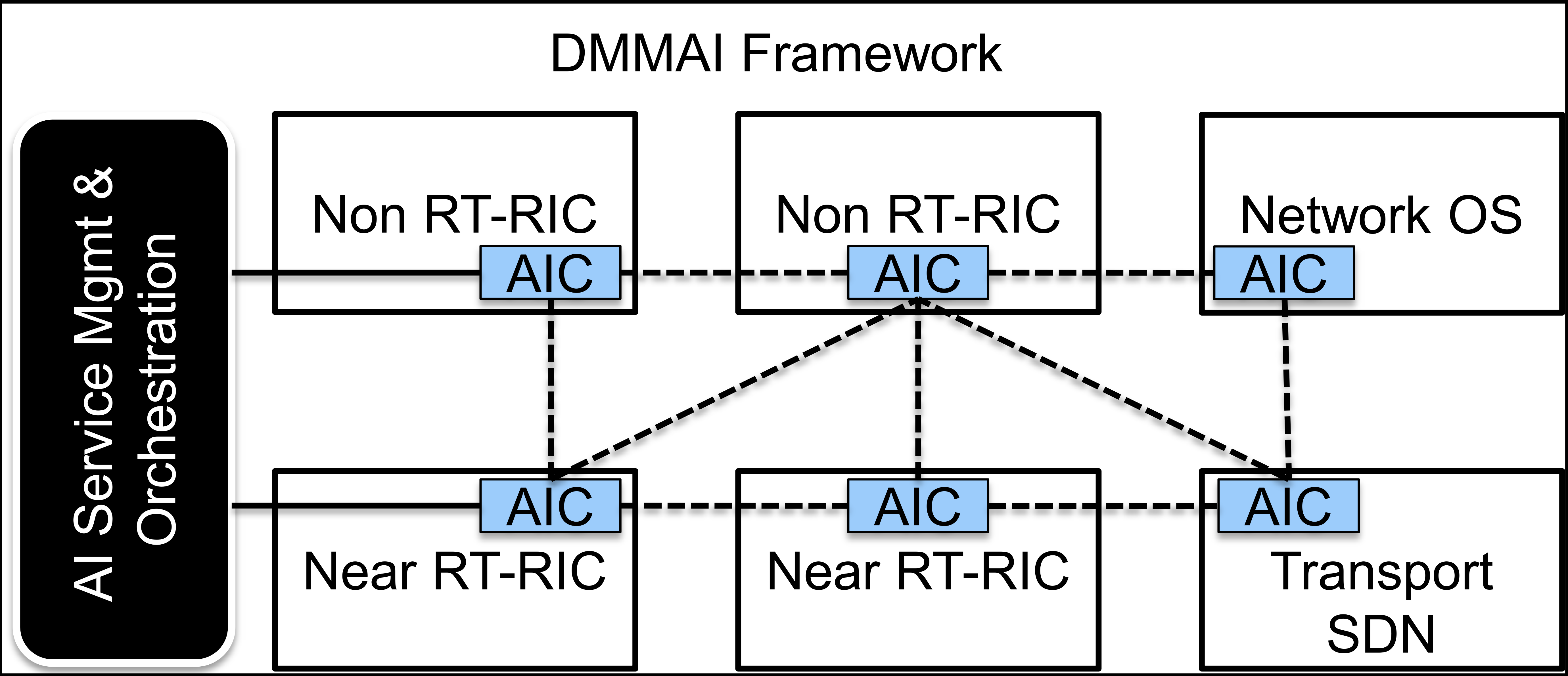}
    \vspace{-3mm}
    \caption{\acl{dmmai} Framework consists of an AI control toolbox that facilitates end-to-end AI services.}
    \vspace{-3mm}
    \label{fig:dmmai_framework}
\end{figure}

Our goal is to propose such a framework and enable the creation of intelligent, zero-touch networks that can configure, monitor, and repair themselves at a large scale. 
The core of our research revolves around understanding the dynamics of multi-network orchestration, achieving a seamless integration of \ac{ai}-driven controls across radio and optical fiber networks. The potential impact of this work is amplified by the current lack of a unified framework or standardized practices for such coordination. This paper proposes a \ac{dmmai} framework for designing mobile networks that seamlessly integrate \ac{ai}, enabling a holistic evaluation of \ac{ai}-driven control mechanisms. The proposed solution is the first step in developing a reference framework for \ac{ai} that will pave the way toward global validation and standardization of \ac{ai} approaches in large-scale 6G networks. It will enable the development of reference use cases, data acquisition, and generation methods, data and model repositories, as well as curated training and evaluation data.

\section{DMMAI Framework}
We have created the \ac{dmmai} framework to tackle the complexities of embedding \ac{ai}/\ac{ml} technologies into network architectures and to facilitate cohesive interaction across all network elements in both the radio and optical domains. It is designed to form a unified `multi-network' approach. Our framework is also meant to serve as a foundational blueprint, providing a detailed yet adaptable structure for further development and enhancements. While it outlines the essential components and functionalities needed for effective \ac{ai}/\ac{ml} integration, it also maintains the flexibility to accommodate future expansions and technological evolutions, ensuring its applicability and relevance in the dynamic landscape of network technology. To achieve this, we will leverage the O-RAN architecture, an open-source platform enabling multi-party applications (xApps/rApps) within a \ac{ric}. Our primary objective is to examine the practical applications of these control mechanisms and their potential for 6G networks. 

Figure~\ref{fig:dmmai_framework} depicts the overall architecture of our proposed framework. It is composed of an \ac{aic} and a cross-controller network architecture. The \ac{aic} resides in various software defined networking environments including: near-\ac{rt} \ac{ric}, non-\ac{rt} \ac{ric}, transport \ac{sdn}, and \ac{nos}. In addition, an \ac{ai} service management and orchestration platform facilitates end-to-end \ac{ai} services within the network. 

The \ac{aic} plays a pivotal role in enhancing the \ac{dmmai} framework's effectiveness and versatility. By decoupling various functionalities, it facilitates a powerful framework structure where \ac{ai} controls can be integrated with different network nodes. This integration supports both \ac{ns} and \ac{ew} communication flows, laying a foundation for the development and implementation of advanced algorithms. 

One of the key advantages of this approach is the facilitation of group or swarm intelligence. This concept involves the coordinated actions of various network nodes, where decision-making and operational strategies are informed and enhanced by shared data insights. The simplicity and modularity of the \ac{dmmai} framework simplify the integration and testing of different \ac{ai}-based platforms, while also allowing for the incorporation of diverse technological solutions. This flexibility ensures that the framework remains adaptable and scalable, capable of evolving with technological advancements and emerging network needs. The \ac{dmmai} framework's design thus offers a unique balance – it is detailed enough to provide clear guidance on essential components, yet open-ended to allow for future extensions and enhancements, ensuring its long-term relevance in the dynamic field of network technology.

\begin{figure}
    \centering
    \includegraphics[scale=0.09]{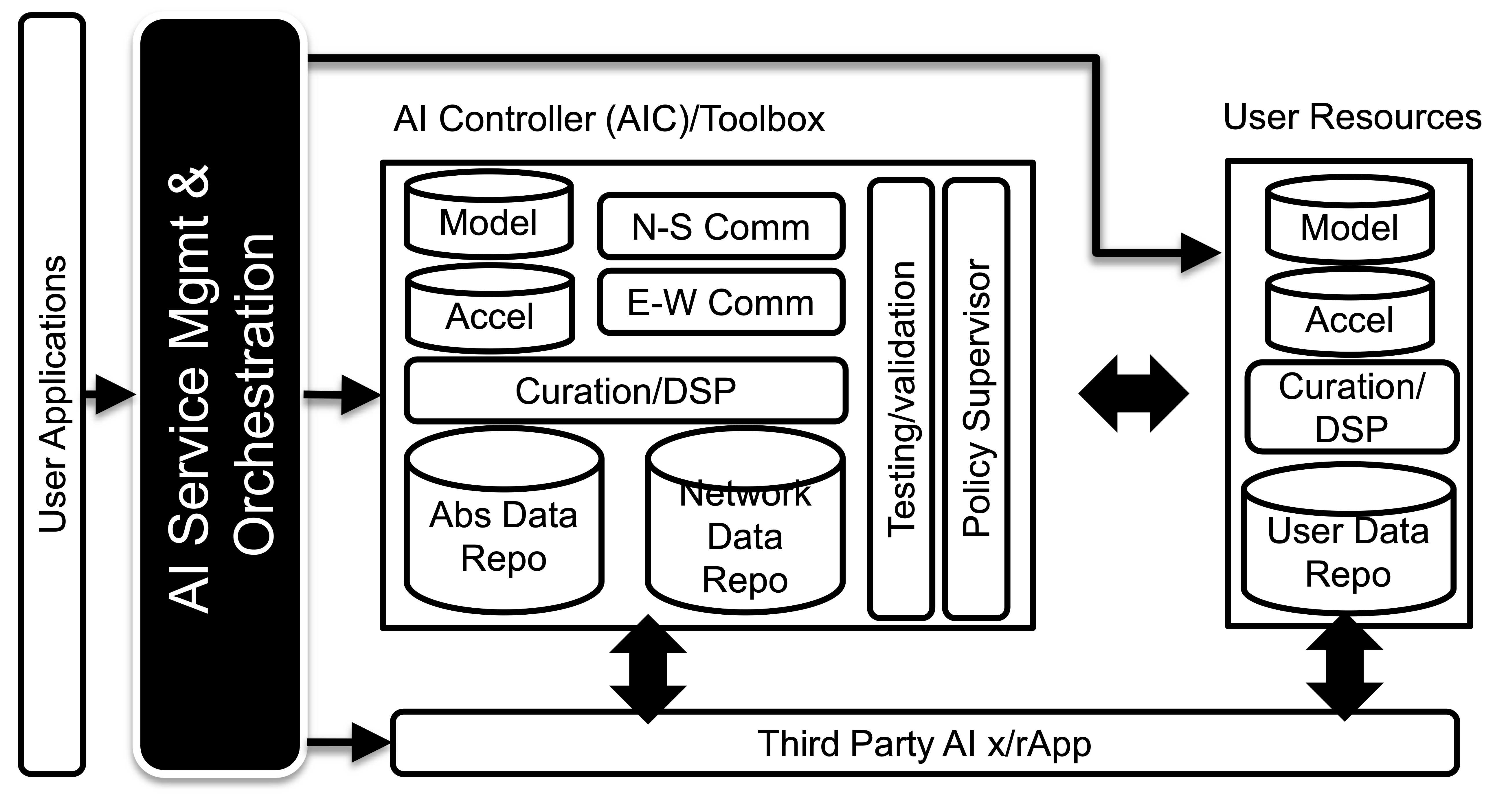}
    \vspace{-3mm}
    \caption{The AI controller or toolbox is composed of a set of repositories, virtual network functions, and interfaces. These interact with both user resources and x/rApps in order to carry out AI-based network functions/control. \vspace{-2em}}
    \label{fig:ai_controller}
\end{figure}

Reference elements of an \ac{aic} are shown in Figure~\ref{fig:ai_controller}. These reference elements are not intended to be comprehensive, and the framework can be further customized. The four classes of elements included here are: 1) data management (curation/DSP, abstracted data repository, and network data repository); 2) model management (model repository and accelerators); 3) inter \ac{aic} communication (\ac{ns}, \ac{ew} APIs/protocols); 4) \ac{ai} management (testing/validation controller, policy supervisor). Each of these building blocks interacts with the respective network elements accessible to the controller in which they are embedded. The building blocks may further interact with \ac{ai} x/rApps or external controllers or orchestrators through the respective \acp{ric} as well as user \ac{ai} resources. The network data repository is intended to house the data accumulated from the respective telemetry system in the given network. The abstracted data repository maintains the processed and curated data for application use. Specific \ac{ai} workflows would be implemented by an orchestrator using these tools within the network. User/application resources include similar data and model management tools for the user applications. The \ac{aic} can be used to configure the network in order to best support different user \ac{ai} applications depending on the network application domain. As an example, x/rApps might be deployed that would utilize an \ac{aic} to optimize a private O-RAN network to support \ac{ai}-driven automation in an industrial factory environment. The orchestration of such applications would be managed through the orchestration system, of which the \ac{ai} service management aspects are modeled as the \ac{ai} Service Management and Orchestration element. This would include a northbound \ac{api} for user applications.

The specification of the different \ac{aic} elements depends on the networks and nodes in which the \ac{aic} is embedded, which is network technology dependent (e.g., radio vs optical). For example, the radio \ac{aic} can leverage O-RAN, whereas the optical \ac{aic} can leverage \ac{onos}. The technologies that they could control range from radio beam forming and spectrum controls to optical signal routing and power management.

\vspace{-0.5em}
\section{Conclusion}
This paper presents a vision for the evolution of 6G communication networks through the \ac{dmmai} framework. Moving beyond traditional data delivery, our approach focuses on developing networks into decentralized intelligence and response systems, essential for meeting the complex demands of future communications. Our work lays out a foundational strategy for AI-driven 6G networks, encapsulating the need for intelligent, flexible, and secure network systems. The \ac{dmmai} framework is designed to seamlessly integrate \ac{ai}/\ac{ml} technologies into network architectures across both radio and optical domains. Its structure is both detailed for immediate implementation and adaptable for future enhancements. 


\begin{acronym}
  \acro{api}[API]{Application Programming Interface}
  \acro{nfv}[NFV]{Network Function Virtualization}
  \acro{sba}[SBA]{Software-Based Architecture}
  \acro{mac}[MAC]{Media Access Control}
  \acro{phy}[PHY]{Physical}
  \acro{ai}[AI]{Artificial Intelligence}
  \acro{aic}[AIC]{\ac{ai} Controller}
  \acro{dmmai}[DMMAI]{Decentralized Multi-party, Multi-network AI}
  \acro{sdn}[SDN]{Software-Defined Networking}
  \acro{onos}[ONOS]{Open Network Operating System}
  \acro{nos}[NOS]{Network Operating System}
  \acro{qot}[QoT]{Quality of Transmission}
  \acro{tip}[TIP]{Telecom Infra Project}
  \acro{ml}[ML]{Machine Learning}
  \acro{ran}[RAN]{Radio Access Network}
  \acro{rt}[RT]{Real Time}
  \acro{ric}[RIC]{\acs{ran} Intelligent Controller}
  \acro{bs}[BS]{Base Station}
  \acro{ue}[UE]{User Equipment}
  \acro{du}[DU]{Distributed Unit}
  \acro{ru}[RU]{Radio Unit}
  \acro{dns}[DNS]{Domain Name System}
  \acro{cu}[CU]{Centralized Unit}
  \acro{zsm}[ZSM]{Zero touch network and service management}
  \acro{pai}[PAI]{Pervasive \ac{ai}}
  \acro{uav}[UAV]{Unmanned Aerial Vehicle}
  \acro{meao}[MEAO]{Multi-Access Edge Computing Application Orchestrator}
  \acro{mec}[MEC]{Multi-Access Edge Computing}
  \acro{mno}[MNO]{Mobile Network Operator}
  \acro{aiaas}[AIaaS]{\ac{ai} as a Service}
  \acro{ns}[N-S]{North-South}
  \acro{ew}[E-W]{East-West}
\end{acronym}

\section*{Acknowledgment}

This work was supported by Horizon Europe SNS Grant 101139194.
Some elements of this work were supported by the Commonwealth Cyber Initiative (CCI), and NSF grants CNS-1827923, EEC-2133516, CNS-2112562, and OAC-2029295.




%
\bibliographystyle{IEEEtran}
\bibliography{main.bib}



\end{document}